\documentclass[prl,aps,twocolumn,showpacs,amsmath,amssymb,floatfix,nobibnotes]{revtex4-2}

\usepackage{comment}
\usepackage{hyperref}
\usepackage{xcolor}
\usepackage{soul}
\usepackage{dsfont}
\usepackage[T1]{fontenc}
\usepackage{physics}
\usepackage{multirow}
\usepackage{tikz}

\usepackage{ulem}

\renewcommand{\vec}[1]{\boldsymbol #1}

\def\12{\frac{1}{2}}

\usepackage{makerobust}
\usepackage{xcolor}
\newcommand{\makeauthor}[2]{\newcommand{#1}[1]{{%
 \sffamily\color{#2}{%
 \bfseries\begingroup\escapechar=-1\edef\x{\endgroup\string#1}\x:%
 } ##1}}%
 \MakeRobustCommand#1}
\makeauthor{\km}{blue}
\makeauthor{\jb}{purple}

\begin{document}
\title{Dynamical qubit cooling via quasi-particle transport}

\author{Kyle Monkman}
\thanks{These authors contributed equally to this work.}
\author{Jasmin Bedow}
\thanks{These authors contributed equally to this work.}

\affiliation{Department of Physics and Astronomy, and Quantum Matter Institute, University of British Columbia, Vancouver, British Columbia V6T 1Z4, Canada}
\date{\today}

\begin{abstract}
Cooling a quantum state presents a challenge as the external control in active protocols can induce detrimental heating in the system. Passive cooling processes offer a valuable alternative: Here, we present a simple quench protocol to cool a set of qubits with access to an ancillary spin chain. We show that excitations can be transported out of the qubits as quasi-particles, taking the form of magnons in a ferromagnetic ancilla and triplons in a dimerized antiferromagnetic ancilla. This quasi-particle transport mechanism enables cooling towards both individual and highly entangled two-qubit target states. We demonstrate that this cooling mechanism remains robust starting from thermal states and with finite quench ramp times.
\end{abstract}
\maketitle

{\it Introduction.~}
Cooling a set of qubits is a critically important task for quantum state initialization. To initialize a target state, its occupation needs to be increased by removing excitations and the associated entropy from the system. This removal process must be implemented carefully to avoid causing additional heating. 

The most well-established cooling protocol is known as algorithmic cooling \cite{AlgLimits,AlgnOthers,Alg2015,AlgCorr,AlgQuantum,AlgPRA,AlgCorrEnh,AlgDyn,AlgQuasiPart}. These protocols utilize a unitary gate description to transport entropy from the system qubits to a set of ancillary qubits. 
Further, to cool complex many-body states, physics-inspired algorithms have been developed to target quasi-particle excitations and remove them using unitary gates \cite{AlgQuasiPart,Google}. Recently, this approach has been demonstrated successfully on two silicon spin qubits operating above 1 Kelvin \cite{spinqubitinit2024}.

In analog computers \cite{Analog1,Bloch2012,Blatt2012,Browaeys2020, PRXQuantum.2.017003}, which have a physical Hamiltonian description, these algorithmic cooling protocols are not easily implementable. Instead of using a complex sequence of pulses to implement the gate-based protocols in digital computers, here one can use the natural physical dynamics to remove the excitations about a many-body state and transport them into the ancilla. For this purpose, several dynamical protocols have been proposed to achieve this shift of the entropy to the ancilla \cite{MBcooling2009,MBcoolingDisentangling,MBcoolingGreiner,MBcoolingPrethermal,MBcoolngZaletel}.

In both approaches, a critical challenge is the initialization of a low-entropy ancilla state capable of absorbing the system qubits' entropy. To achieve this, one strategy is to thermalize the ancilla with a larger effective magnetic field compared to the main qubit system. However, varying effective magnetic fields across subsystems is not a practical strategy for all quantum computing platforms. Changing the local magnetic field for individual qubits can carry a significant overhead \cite{LocalTuning_ColdAtom,LocalTuning_ColdAtoms2,LocalTuning_Rydberg}, induce crosstalk between neighboring sites \cite{Crosstalk_Rydberg}, and in some cases is unavailable all-together \cite{FixedfrequencySCqubits}.

To address this challenge, we initialize low-entropy thermal states using the emergent ordering of many-body states. For instance, we utilize the natural ordering of spins in ferromagnetic and antiferromagnetic chains which reduces the overall entropy. Using this principle, we prepare low-entropy ancilla states for the purpose of cooling the system qubits. This ancilla has lower entropy than the qubits even while maintaining uniform magnetic field. Unlike previous studies designed to cool a many-body state, this takes the reverse approach of using the ordering of a many-body state to cool qubits of interest.

Using the initialized ancilla state, the goal is to increase the occupation of a target state of the system qubits. To do this, we use the quasi-particle bands in the ancillary many-body state to remove entropy from the system qubits. First, we cool a single qubit into a spin-down state by transporting any spin-up excitations into a magnon \cite{magnonMediatedGates,magnonRoadmap2024,PRXQuantum.2.040314, PhysRevX.3.041023} band of a ferromagnetic chain. We then cool a pair of qubits into a highly-entangled singlet state by coupling them to a triplonic quasi-particle band \cite{Triplon2,Triplon3,Triplon4,Triplon5,Triplon6,Triplon7,Triplon8} in a dimerized antiferromagnet \cite{Sachdev1990}.  We validate that this transport process increases the target state population for both pure and thermal initial states. This protocol is robust in the experimentally relevant case where the couplings between the system and ancilla have a finite ramp time. Ultimately, our approach represents a low-overhead cooling protocol using the natural ordering of many-body states as a cooling advantage.

{\it Quasi-particle transport mechanism.~}
We consider first a ferromagnetic Heisenberg exchange Hamiltonian for a one-dimensional spin $1/2$ chain given by
\begin{eqnarray}
    \mathcal{H}_{\text{F},0}=-J\sum_{j=2}^{N-1} \vec{S}_j \cdot \vec{S}_{j+1}+B\sum_{j=1}^N S_j^z \; ,
\end{eqnarray}
which acts on a $2^N$ dimensional Hilbert space of $N$ spin $1/2$ states. The exchange coupling $J$ acts only between nearest-neighbor sites starting from the second site $j=2$. The magnetic field $B>0$ acts uniformly on all spins. We consider the first spin $j=1$ as the system qubit, which is initially decoupled from the other magnetic moments $j \geq 2$. The ground state of $\mathcal{H}_{\text{F},0}$ is the decoupled state $\ket{q} \otimes \ket{\psi_G^a}$ where $\ket{q}=\ket{\downarrow}$ is for the system qubit and  $\ket{\psi_G^a}=\ket{\downarrow \downarrow \dots \downarrow}$ is the ancilla ground state.

\begin{figure}
    \centering
    \includegraphics[width=\linewidth]{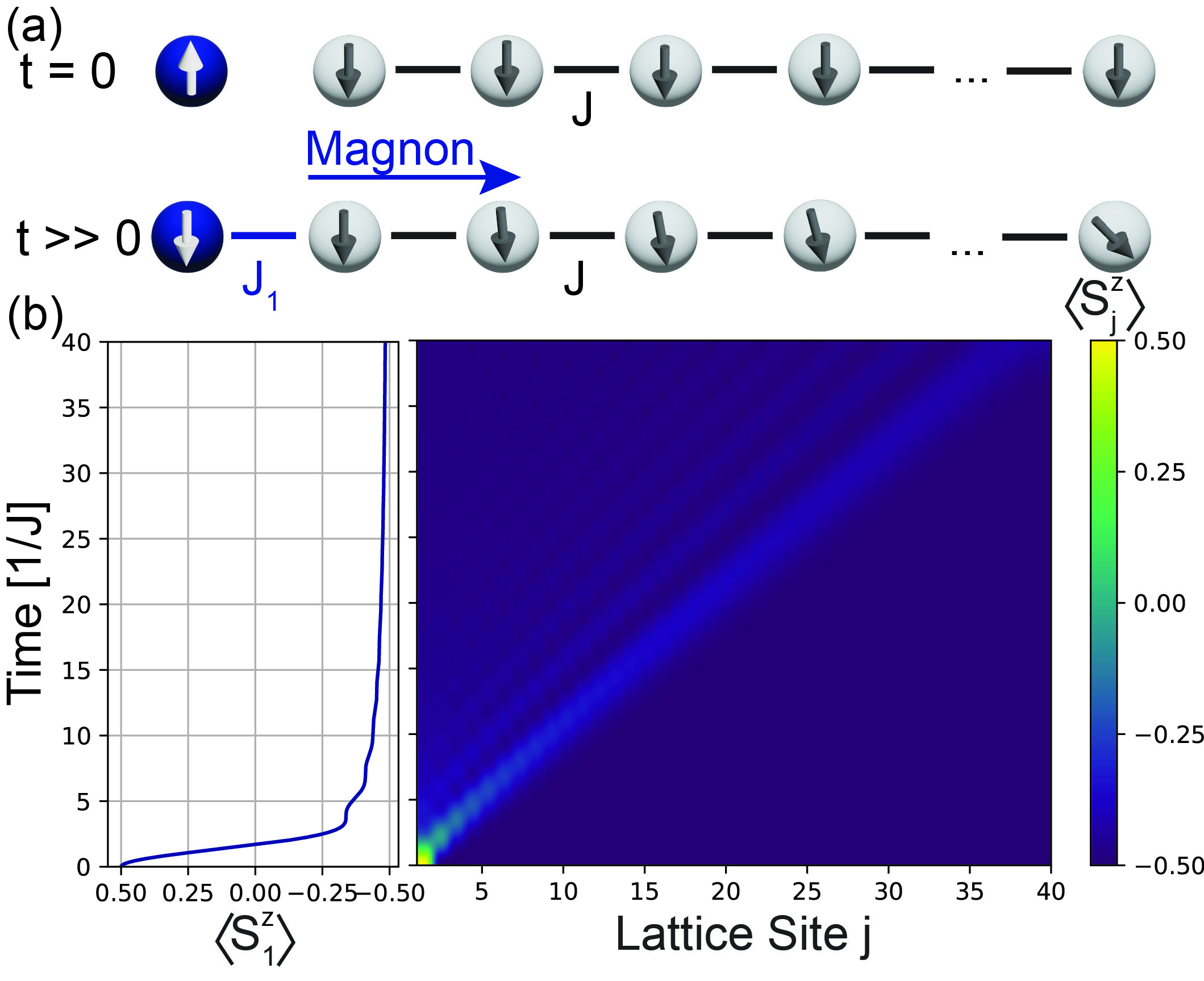}
    \caption{(a) Schematic of the magnonic transport mechanism: The excited state in the qubit creates magnonic excitations in the ferromagnetic ancilla.  (b) Light cone diagram of $\langle S_j^z \rangle$. The magnon travels through the ancilla, removing the excitation from the qubit, with the inset on the left showing $\langle S_1^z \rangle$ as a function of time. As $\langle S_1^z \rangle \rightarrow -0.5$, the magnonic excitation is fully within the ancilla and the qubit is reset to a spin-down state. Parameters are $J = 1, J_1 = 1, B = 0.5$, $N=40$.}
    \label{fig:Fig1}
\end{figure}

We time-evolve this system with the Hamiltonian
\begin{eqnarray}
    \mathcal{H}_{\text{F},1}=-J_1\vec{S}_1 \cdot \vec{S}_{2}+\mathcal{H}_{\text{F},0}
\end{eqnarray}
and the associated time evolution operator $\exp(-\mathrm{i}\mathcal{H}_{\text{F},1}t)$ for $t>0$. Physically, this corresponds to a quench of the exchange parameter $J_1$ between the qubit and the $j=2$ spin of the ancilla chain. Starting from a system initialized in the excited state $\ket{\uparrow} \otimes \ket{\psi_G^a}$, the transport process is shown schematically in Fig.~\ref{fig:Fig1}(a) for $t=0$ (upper plot) and $t\gg0$ (lower plot). In Fig.~\ref{fig:Fig1}(b) (left), we plot $\langle S_1^z \rangle(t)$ for a system of size $N=40$. The expectation value $\langle S_1^z \rangle(t)$ approaches $-1/2$, indicating that the qubit approaches the $\ket{\downarrow}$ state, effectively resetting the qubit.

This transport process is understood by introducing an effective quasi-particle magnon description. We define quasi-particle operators $a_j$ using $S_j^+ = a_j^\dag$ and $S_j^z=\left(a_j^\dag a_j -\frac{1}{2}\right)$. These operators $a_j$ satisfy the commutation relations for hardcore bosons $[a_i, a_j^\dag]=\delta_{i,j}(1-2a_i^\dag a_i)$, $[a_i,a_j]=0$ and $(a_i)^2=0$. Using these operators, the dynamical ferromagnetic interaction becomes  
\begin{align}
    &\mathcal{H}_{\text{F},1}= B\sum_{j=1}^N \left( a_j^\dag a_j -\frac{1}{2}\right)\nonumber \\ 
    &-\sum_{j=1}^{N-1} J_j \left( (a_j^\dag a_j -\frac{1}{2}) (a_{j+1}^\dag a_{j+1} -\frac{1}{2}) + \frac{1}{2} (a_j^\dag a_{j+1}+\text{h.c.})\right)
\end{align}
where $J_j=J$ for $j\geq 2$. Here, as $[\mathcal{H}_{\text{F},1}, \hat{N}]=0$, with $\hat{N}=\sum_{j=1}^N a_j^\dag a_j =\sum_{j=1}^N (S_j^z+1/2)$ the total number of magnons and the total spin-$z$ component is conserved in the dynamics.

The initialized excited state $\ket{\uparrow} \otimes \ket{\psi_G^a}=a_1^\dag \ket{\downarrow} \otimes \ket{\downarrow \downarrow \dots \downarrow}$ describes a single magnon in the qubit. Due to the conservation of $\hat{N}$, there remains only a single magnonic excitation as time evolves via $\mathcal{H}_{\text{F},1}$. Therefore, the interaction term $a_j^\dag a_j a_{j+1}^\dag a_{j+1}$ has no effect on the time evolution. In Fig.~\ref{fig:Fig1}(b) (right), we plot $\langle S_j^z\rangle=\langle (a_j^\dag a_j-1/2)\rangle$ as a color plot in position and time. As shown, the magnon occupation travels ballistically \cite{ballistic} away from the qubit, creating a linear light cone. Thus, the removal of the magnonic quasi-particle constitutes a resetting of the qubit to the $\ket{\downarrow}$ state.

\begin{figure}
    \centering
    \includegraphics[width=\linewidth]{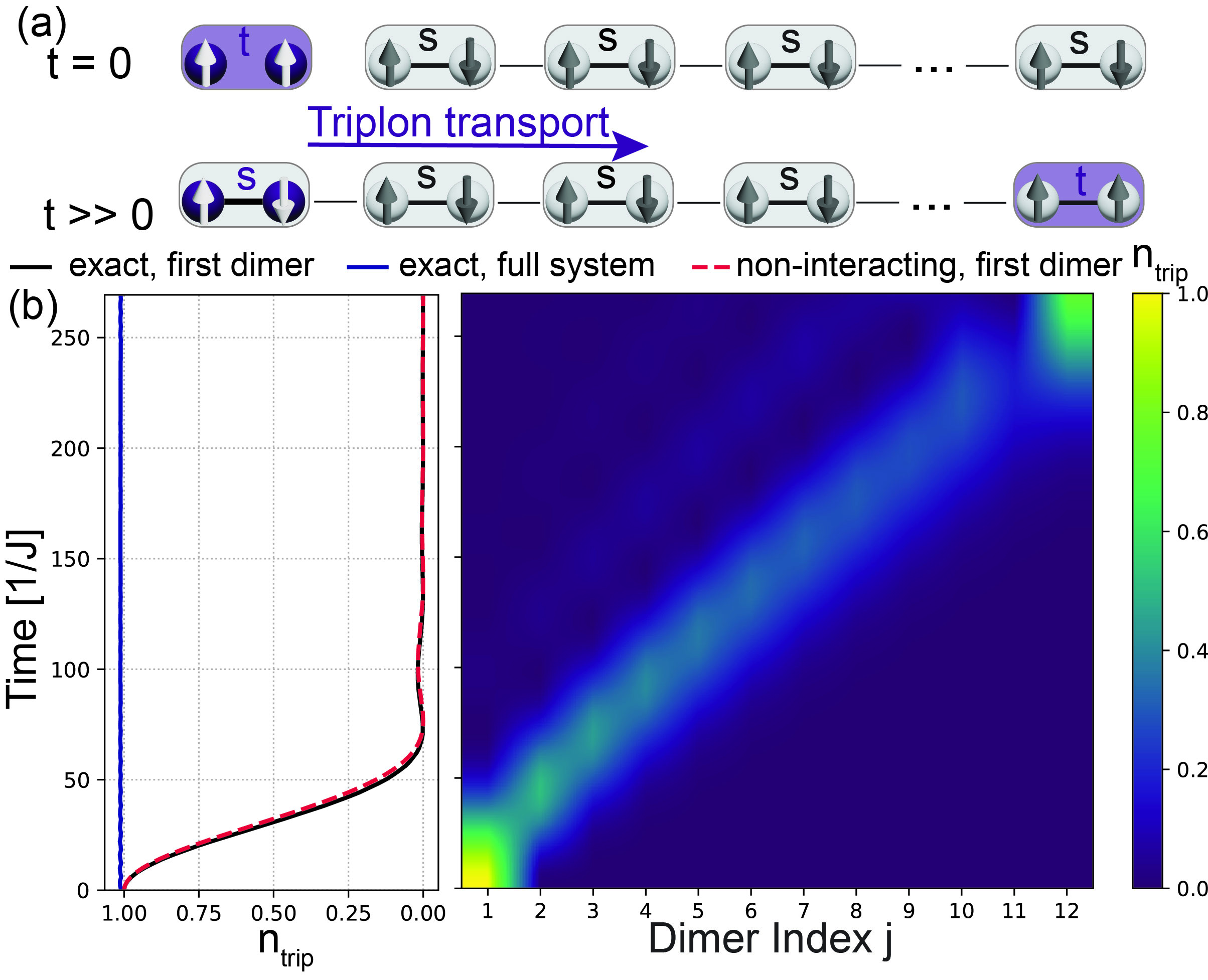}
    \caption{(a) Schematic of the triplonic transport mechanism: The excited states in the qubits create triplonic excitations in the dimerized antiferromagnetic ancilla. (b) Triplon occupation $\sum_\sigma \langle t_{j,\sigma}^\dag t_{j,\sigma} \rangle$ for the qubit dimer (left) and for the entire system (right) as a function of time for parameters $J=1$, $J'=0.1$, $B=0.1$ from Eq.~\eqref{eq:hamAF0}, and $J_1=1$, $J_1'=0.1$ from Eq.~\eqref{eq:hamAF1} for a system of 12 unit cells.  Shown is the exact dynamics and the quasi-particle approximation. Exact dynamics were simulated using a first-order ODE solver by \textit{QuSpin} \cite{quspin1,quspin2}.}
    \label{fig:Fig2}
\end{figure}

Second, we consider a dimerized antiferromagnetic spin 1/2 Heisenberg model with a two-site unit cell, described by
\begin{align}
    \mathcal{H}_{\text{AF},0} &= J\sum_{j=2}^{N_u} \vec{S}_{j,A} \cdot \vec{S}_{j,B}+J'\sum_{j=2}^{N_u-1} \vec{S}_{j,B} \cdot \vec{S}_{j+1,A} \nonumber \\
    &+ B\sum_{j=1}^{N_u} (S_{j,A}^z+S_{j,B}^z) \; ,
    \label{eq:hamAF0}
\end{align}
with $N_u$ unit cells, total spin sites $N=2N_u$ and a total Hilbert space of dimension $2^N=4^{N_u}$. Here, the spins $\vec{S}_{j,A}$ and $\vec{S}_{j,B}$ refer to the two spins in each unit cell. The first two spins $S_{1,A}$ and $S_{1,B}$ are the two system qubits, which are initially completely decoupled. The ground state of $\mathcal{H}_{\text{AF},0}$ is also a decoupled state $\ket{\downarrow \downarrow} \otimes \ket{\psi_G^a}$, where $\ket{\psi_G^a}$ is the ancilla ground state.

We then time-evolve with the Hamiltonian
\begin{equation}
    \mathcal{H}_{\text{AF},1}= J_1 \vec{S}_{1,A} \cdot \vec{S}_{1,B}+ J_1'\vec{S}_{1,B} \cdot \vec{S}_{2,A}+\mathcal{H}_{\text{AF},0} \; .
    \label{eq:hamAF1}
\end{equation}
Analogous to the ferromagnet case, the magnetic field $B$ is uniform throughout the system and unchanged in both $\mathcal{H}_{\text{AF},0}$ and $\mathcal{H}_{\text{AF},1}$. For $\mathcal{H}_{\text{AF},1}$, we add a coupling within the two system qubits and between the system and the ancilla, respectively. 

To understand the occurring dynamics, we utilize the bond operator representation for singlets and triplets on a dimerized antiferromagnet, as introduced in Ref. \cite{Sachdev1990}. We define for each dimer singlet $\ket{s_j}$ and triplet states $\ket{t_{j,\sigma}}$ for $\sigma=x,y,z$ as
\begin{equation}
\label{singletTriplet}
    \begin{aligned}
        \ket{s_j} &= s_j^\dagger \ket{\mathrm{vac}} = \frac{1}{\sqrt{2}} (\ket{\uparrow \downarrow}_j - \ket{\downarrow \uparrow}_j) \\
        \ket{t_{j,x}} &= t_{j,x}^\dagger \ket{\mathrm{vac}} = -\frac{1}{\sqrt{2}} (\ket{\uparrow\uparrow}_j - \ket{\downarrow \downarrow}_j) \\
        \ket{t_{j,y}} &= t_{j,y}^\dagger \ket{\mathrm{vac}} = \frac{\mathrm{i}}{\sqrt{2}} (\ket{\uparrow\uparrow}_j + \ket{\downarrow \downarrow}_j) \\
        \ket{t_{j,z}} &= t_{j,z}^\dagger \ket{\mathrm{vac}} = \frac{1}{\sqrt{2}} (\ket{\uparrow \downarrow}_j + \ket{\downarrow \uparrow}_j) \; ,
    \end{aligned}
\end{equation}
where $\ket{\mathrm{vac}}$ is the vacuum state. These operators satisfy the physical constraint that for all $j$ only one singlet and/or triplet can be present per dimer, yielding $s_j^\dag s_j +\sum_{\sigma=x,y,z} t_{j,\sigma}^\dag t_{j,\sigma}=1$. Furthermore, they satisfy bosonic commutation relations $[s_i,s_j^\dag]=\delta_{i,j}$, $[t_{i,\sigma},t_{j,\sigma'}^\dag]=\delta_{i,j} \delta_{\sigma,\sigma'}$ and $[s_i,t_{j,\sigma}^\dag]=0$.

Then, the terms of the Hamiltonians $\mathcal{H}_{\text{AF},0}$, $\mathcal{H}_{\text{AF},1}$ can be rewritten as 
\begin{align}
	\label{spins2bonds}
	\mathbf{S}_{j,A} \cdot \mathbf{S}_{j,B} &= \frac{1}{4} t_{j,\sigma}^\dagger t_{j,\sigma} - \frac{3}{4} s_j^\dagger s_j  \\ 
	\begin{split}
		\mathbf{S}_{j,B} \cdot \mathbf{S}_{j+1,A} &= \frac{1}{4} (-s_j^\dagger t_{j,\sigma} - t_{j,\sigma}^\dagger s_j - \mathrm{i} \varepsilon_{\sigma\beta\gamma} t_{j,\beta}^\dagger t_{j,\gamma}) \\ \times &( s_{j+1}^\dagger t_{j+1,\sigma} + t_{j+1,\sigma}^\dagger s_{j+1} - \mathrm{i} \varepsilon_{\sigma\beta\gamma} t_{j+1,\beta}^\dagger t_{j+1,\gamma})
	\end{split} \label{eq:intertriplon} \\
	\begin{split}
		(S_{j,A}^z+S_{j,B}^z) &= \mathrm{i}( t_{j,y}^\dagger t_{j,x}-t_{j,x}^\dagger t_{j,y}).
	\end{split} 
\end{align}
Here, we consider the regime where $J \gg J',B$ so that the internal exchange terms $\mathbf{S}_{j,A} \cdot \mathbf{S}_{j,B}$ described above are the largest terms in $\mathcal{H}_{\text{AF},0}$. In this case, we have the approximate ancilla ground state $\ket{\psi_G^a} \approx \ket{s_2, \dots s_{N_u}}=s_2^\dag, \dots s_{N_u}^\dag |\text{vac}\rangle$. 

Then, we define `triplonic' quasi-particle excitations $T_{j,\sigma}^\dag = t_{j,\sigma}^\dag s_j $ as excitations on the ancilla ground state $T_{j,\sigma}^\dag \ket{\psi_G^a}$. Although the number of triplons is not strictly conserved under $\mathcal{H}_{\text{AF},1}$, they cost an energy of approximately $J$ to create. This makes them approximately conserved purely based on energy constraints. To study this conservation, we consider the operators
\begin{eqnarray}
    n_{j,\text{trip}}&=&\sum_\sigma T_{j,\sigma}^\dag T_{j,\sigma}=\sum_\sigma t_{j,\sigma}^\dag t_{j,\sigma}=1-s_j^\dag s_j \nonumber \\
    N_{\text{trip}}&=&\sum_j n_{j,\text{trip}} \; ,
\end{eqnarray}
where the second equality comes from the bosonic commutation relations.

We define our transport mechanism based on this quasi-particle description. We start in an initial state $\ket{t_{1,x}} \otimes \ket{\psi_G^a}$ where $\ket{\psi_G^a}$ is the ancilla ground state of $\mathcal{H}_{\text{AF},0}$, depicted at $t=0$ in Fig. \ref{fig:Fig2}a. As this state evolves governed by $\mathcal{H}_{\text{AF},1}$, the triplonic excitation in the qubits is transported into the ancilla, also shown in Fig. \ref{fig:Fig2}(a) for $t\gg0$. We simulate this transport process for $J=1$, $J'=0.1$ and $B=0.1$ We plot $\langle n_{j,\text{trip}} \rangle$ in Fig. \ref{fig:Fig2}b (right), showing a linear transport of the triplon away from the qubits. As the triplon leaves the qubits, the occupation $\langle n_{1,\text{trip}} \rangle \rightarrow 0$ shown in Fig.\ref{fig:Fig2}(b) (left). Equivalently, the occupation $\langle s_1^\dag s_1 \rangle$ approaches 1, demonstrating a reset of the qubit pair into a singlet state. Due to the rotational symmetry of $\langle n_{1,\text{trip}} \rangle$, the dynamics of this operator are equivalent for any initial starting state $\ket{t_{1,\sigma}} \otimes \ket{\psi_G^a}$.

This transport process is closely described by the non-interacting triplon description since $\langle N_{\text{trip}} \rangle$ remains close to 1, shown in Fig. \ref{fig:Fig2}(b) (left). Neglecting the interacting triplonic terms in Eq.\eqref{eq:intertriplon}, we have the approximate description
\begin{align}
    &\mathcal{H}_{\text{AF},1} \approx -\mathrm{i}B \sum_{j=1}^{N_u} (T_{j,x}^\dag T_{j,y}-T_{j,y}^\dag T_{j,x}) \nonumber \\ &+\sum_{j=1}^{N_u} J_j \left(T_{j,\sigma}^\dag T_{j,\sigma}-\frac{3}{4}\right) - \frac{1}{4} \sum_{j=1}^{N_u-1}J_j'( T_{j,\alpha}^\dagger T_{j+1,\alpha} + \text{h.c.}) ,
\end{align} 
with an ancilla initialized exactly in $\ket{s_2,\dots,s_{N_u}}$. We plot the reset parameters $\langle n_{1,\text{trip}} \rangle$ in Fig \ref{fig:Fig2}b (left), showing the close agreement with the exact dynamics.

\begin{figure}[t]
    \centering
    \includegraphics[width=\linewidth]{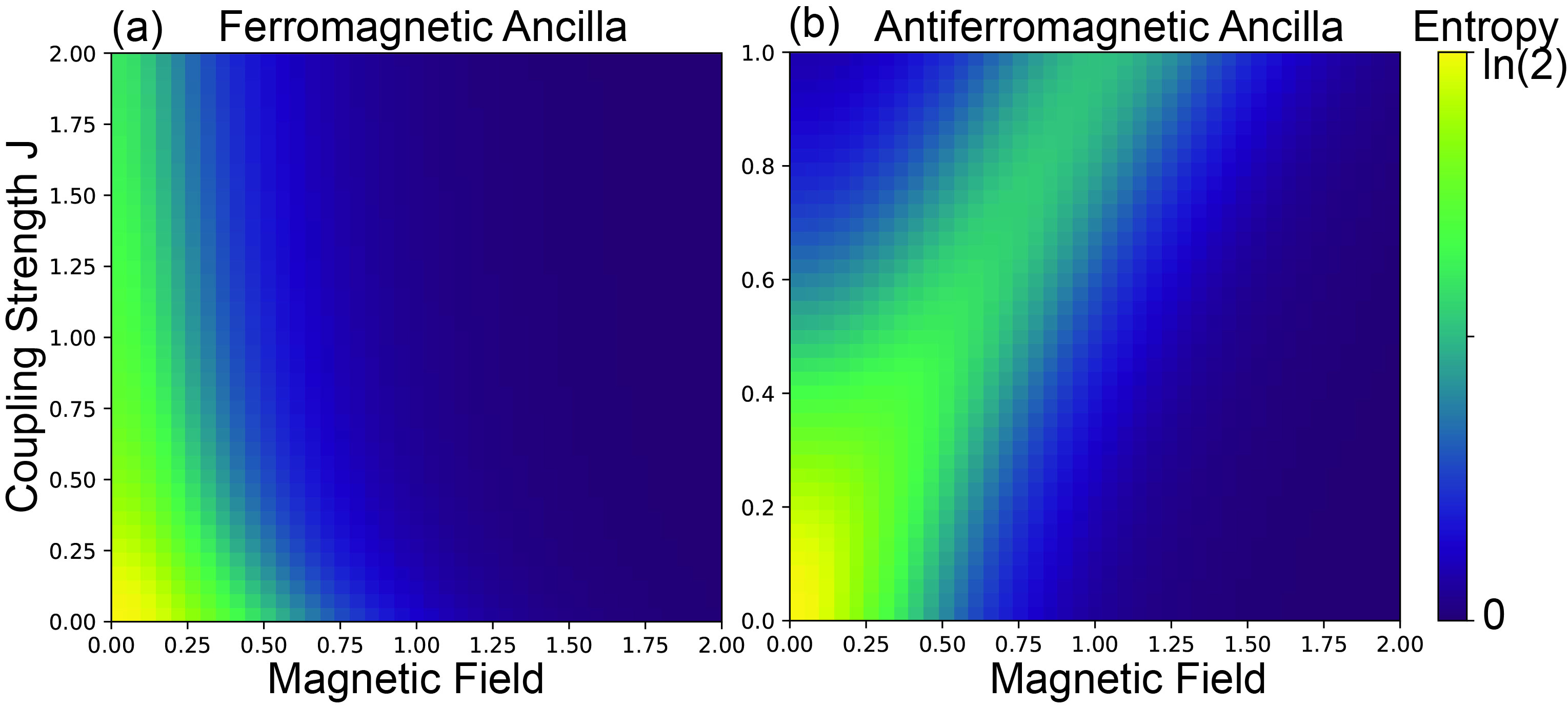}
    \caption{Average entropy of the initial thermal state for the system and ancilla states in the (a) ferromagnetic and (b) antiferromagnetic case as a function of magnetic field $T$ and exchange coupling $J$. Parameters are $\beta =4$ for (a), and $\beta=5, J' = 0.1J$ for (b).}
    \label{fig:Fig3}
\end{figure}

{\it Cooling thermal states.~} The quasi-particle process described above can be extended to a cooling protocol if the ancilla has a lower average entropy than the qubits. We start our protocol from thermal states of $\mathcal{H}_{\text{F},0}$ and $\mathcal{H}_{\text{AF},0}$. In particular, we look at $\rho_\text{S}=\exp(-\beta \mathcal{H}_{\text{S},0})/Z_{\text{S}}$ for partition functions $Z_{S}$ with $S \in \{ \text{FM}, \text{AF}\}$. Since the qubits and ancilla are initially decoupled we have $\rho_S=\rho_{q,S}\otimes \rho_{a,S}$. Then, we calculate the average entropies $S_{q,S}=-\Tr \rho_{q,S} \log \rho_{q,S} / N_q$ and $S_{a,S}=-\Tr \rho_{a,S} \log \rho_{a,S} / N_a$ with $N_q$ being the number of qubits in the system and $N_a$ the number of spins in the ancilla. 

The average entropies are exactly equal $S_{q,S}=S_{a,S}$ in the case when there are no exchange interactions $J=0$, $J'=0$. Since only the ancilla has interactions in $\mathcal{H}_{S,0}$, its entropy $S_{a,S}$ can be reduced by increasing $J,J'>0$. This is shown in Fig.~\ref{fig:Fig3}, where we present the average entropy in the ancilla for both the ferromagnetic (a) and antiferromagnetic (b) cases as a function of the magnetic field $B$ and coupling strength $J$ with $J'=0.1J$ in (b). In both cases, this showcases the ability to prepare an ancilla state with lower average entropy than the system qubits, while maintaining a uniform magnetic field. Note that for the antiferromagnetic ancilla, the magnetic field needs to be kept under a critical value, above which the ground state is no longer antiferromagnetically correlated.

\begin{figure}
    \centering
    \includegraphics[width=\linewidth]{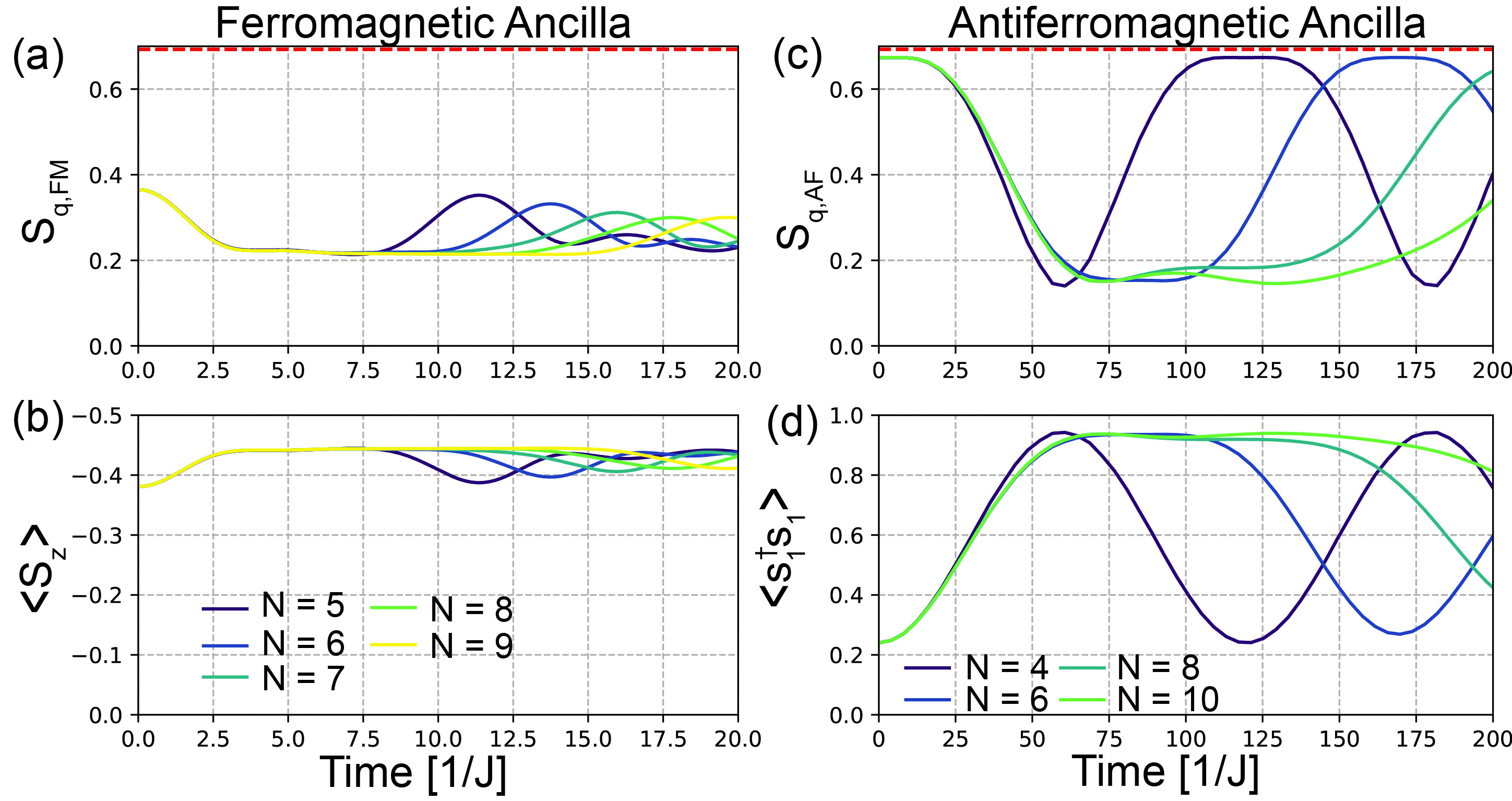}
    \caption{(a) Time-dependent entropy and (b) spin expectation value of the qubit for the ferromagnetic set-up for different numbers of sites $N \in \{5,6,7,8,9\}$ in the entire system. Parameters are $J = 1, J_1 = 1, B = 0.5, \beta = 4$.  (c) Time-dependent entropy and (d) singlet occupation value $\langle s_1^\dag s_1 \rangle=1-\langle n_{1,\text{trip}} \rangle$ of the dimer for the antiferromagnetic set-up for different numbers of sites $N \in \{4,6,8\}$in the entire system. Parameters are $J = 1, J' = 0.1, J_1 = 1, J_1' = 0.1, B=0.1, \beta = 4$.}
    \label{fig:Fig4}
\end{figure}

This low-entropy ancilla can then be used to dynamically cool the qubits via the quasi-particle transport mechanism. We start from the thermal states $\rho_S$ and evolve unitarily via Hamiltonians $\mathcal{H}_{S,1}$. For the ferromagnet $S=\text{F}$, we plot the average qubit entropy $S_{q,\text{F}}$ and the spin-$z$ expectation value $\langle S_1^z \rangle$ in Fig. \ref{fig:Fig4} (a,b). We plot various system sizes $N \in \{5,6,7,8,9\}$ to demonstrate that a plateau arises, as the system size increases. The values approach a cooled qubit steady state with stability time increasing with system size. The antiferromagnet cooling has a similar qualitative process showing increased cooling. For this case, we present the average entropy $S_{q,\text{AF}}$ and the singlet occupation $\langle s_1^\dag s_1 \rangle=1-\langle n_{1,\text{trip}} \rangle$ for the system qubits in Fig. \ref{fig:Fig4} (c,d). Note that in both of these cases, the expectation values plotted in \ref{fig:Fig4} (b,d) are directly proportional to the fidelity of the target state. Hence, we demonstrate a successful lowering of the system qubits' effective temperature in both cases.

Finally, we consider the experimental situation where it takes a ramp time $T_{\text{ramp}}$ to change the system parameters. Then we ramp the changes on and off for a total active quench time of $T$. During the time windows $[0,T_\text{ramp}]$ and $[T-T_\text{ramp},T]$ the parameters are changing linearly as they are turned on and off, respectively. In order to cool the system, the active quench time $T$ is chosen such that it falls during the plateau times, shown in Fig.\ref{fig:Fig4}.

\begin{figure}
    \centering
    \includegraphics[width=\linewidth]{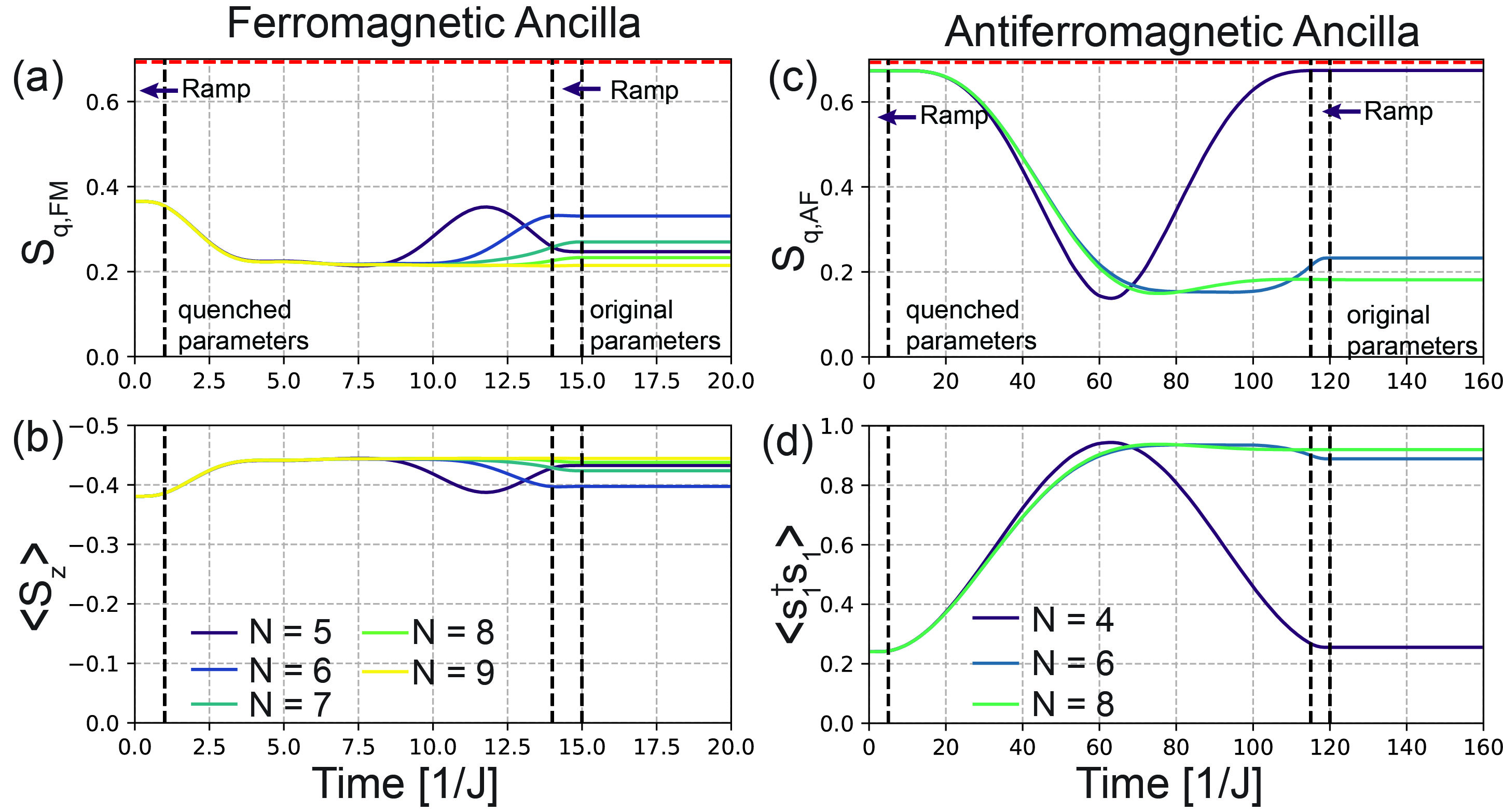}
    \caption{(a) Time-dependent entropy and (b) spin expectation value of the qubit for the ferromagnetic set-up for different lengths $L$ of the ancilla for a ramped quench. Parameters are $J = 1, J_1 = 1, B = 0.5, \beta = 4, T_\text{ramp}=1, T=15$.  (c) Time-dependent entropy and (d) singlet fidelity of the dimer for the antiferromagnetic set-up for different lengths $L$ of the ancilla for a ramped quench. Parameters are $J = 1, J' = 0.1, J_1 = 1, J_1' = 0.1, B=0.1, \beta = 4, T_\text{ramp}=5, T=120$.}
    \label{fig:Fig5}
\end{figure}

We simulate this ramped quench scenario for both the ferromagnetic ancilla and the antiferromagnetic ancilla in Fig.~\ref{fig:Fig5}. To demonstrate selecting the quench time $T$, we fix $T$ and vary system size. We use $T_\text{ramp}=1/J$, $T=15/J$ in the ferromagnetic, and $T_\text{ramp}=5/J$, $T=120/J$ in the antiferromagnetic case. In the ferromagnetic case, shown in Fig.~\ref{fig:Fig5}(a,b), for the system sizes $N=5,6$ the ramp time is too long, beyond the stability time. For $N=7,8,9$, the cooling is effectively implemented after the protocol. The antiferromagnetic case, shown in Fig.~\ref{fig:Fig5}(c,d), $N=4,6$ is too small while $N=8$ has significant stable cooling.  Moreover, we can see that the finite ramp time does not influence the plateaus too much as they remain qualitatively the same as in Fig.~\ref{fig:Fig4}. Importantly, the ramp time cannot be too long, as then the system would start behaving adiabatically.

{\it Discussion.~}
We have demonstrated that quench protocols where qubits are coupled to a correlated set of ancillas is an efficient way of removing unwanted excitations from the qubits. This strategy can be utilized both to purify a single qubit or initialize entangled singlet states in a pair of qubits. We have considered both the case of individual many-body states and thermal states described by density matrices being cooled, and extended the latter to the experimentally realistic ramp protocols. Therein, we considered unitary time-evolution, which is valid when the time associated with the coupling $1/J$ is much smaller than the thermalization time. A natural setting for realizing our proposal is given by quantum dot-based spin qubits \cite{spinqubitinit2024,spinqubits1,spinqubits2}, where the inter-dot hopping parameter can be controlled using gates, allowing for the required changes in the Heisenberg couplings considered here. Having demonstrated that multiple qubits can be initialized into a maximally entangled state, the question naturally arises whether larger numbers of qubits can be initialized into highly entangled states in the same manner. Furthermore, it may be possible to extend this protocol to a broader class of operations as in Ref. \cite{cheung2025spectrally}.

{\it Acknowledgments.~} The authors would like to thank Nitin Kaushal for insightful discussions on triplonic quasi-particles.

\bibliography{bibliography}
\end{document}